\documentclass[prl,twocolumn,showpacs,superscriptaddress,floatfix]{revtex4}
\usepackage{amssymb}
\usepackage{graphicx,amsmath}
\usepackage{bm}
\usepackage{times}
\usepackage{hyperref}

\newcommand{\NLij}{i\kern -0.08em j}

\begin{document}

\title{Consistency of ground state and spectroscopic measurements on flux
qubits}
\author{A.~Izmalkov }
\thanks{deceased}
\affiliation{Institute of Photonic Technology, P.O. Box 100239, D-07702 Jena, Germany}
\author{S. H. W. van der Ploeg}
\affiliation{Institute of Photonic Technology, P.O. Box 100239, D-07702 Jena, Germany}
\author{S. N. Shevchenko}
\affiliation{Institute of Photonic Technology, P.O. Box 100239, D-07702 Jena, Germany}
\affiliation{B. Verkin Institute for Low Temperature Physics and Engineering, 47 Lenin
Ave., 61103, Kharkov, Ukraine}
\author{M.~Grajcar}
\affiliation{Institute of Photonic Technology, P.O. Box 100239, D-07702 Jena, Germany}
\affiliation{Department of Solid State Physics, Comenius University, SK-84248 Bratislava,
Slovakia}
\author{E.~Il'ichev}
\email{evgeni.ilichev@ipht-jena.de}
\affiliation{Institute of Photonic Technology, P.O. Box 100239, D-07702 Jena, Germany}
\author{U.~H\"ubner}
\affiliation{Institute of Photonic Technology, P.O. Box 100239, D-07702 Jena, Germany}
\author{A. N. Omelyanchouk}
\affiliation{B. Verkin Institute for Low Temperature Physics and Engineering, 47 Lenin
Ave., 61103, Kharkov, Ukraine}
\author{H.-G.~Meyer}
\affiliation{Institute of Photonic Technology, P.O. Box 100239, D-07702 Jena, Germany}
\date{\today }

\begin{abstract}
We compare the results of ground state and spectroscopic
measurements carried out on superconducting flux qubits which are
effective two-level quantum systems. For a single qubit and for two
coupled qubits we show excellent agreement between the parameters of
the pseudospin Hamiltonian found using both methods. We argue, that
by making use of the ground state measurements the Hamiltonian of
$N$ coupled flux qubits can be reconstructed as well at temperatures
smaller than the energy level separation. Such a reconstruction of a
many-qubit Hamiltonian can be useful for future quantum information
processing devices.
\end{abstract}

\maketitle

Quantum systems are generally characterized by spectroscopic
measurements: the system is excited by electromagnetic radiation,
with a frequency which matches the level spacing, and the response
of this excitation is detected. On the other hand, quantum theory
predicts that the Hamiltonian of some quantum-mechanical systems can
be completely reconstructed from their ground-state properties. For
instance, quantum mechanical treatment of the ammonia molecule in a
two-level approximation, shows that its ground state contains
information about time-independent Hamiltonian
parameters~\cite{Fey}. Superconducting qubits are also described by
a similar Hamiltonian~\cite{Makhlin01}. They are micrometer-size
quantum systems~\cite{You05} which can be easily accessed by a
macroscopic measuring device. For example, the Hamiltonian
parameters of a superconducting flux qubit~\cite{Mooij99} can be
determined from the measurement of its magnetic susceptibility in
the ground state~\cite{Greenberg02b}. In this Letter we will
demonstrate that for a single and two coupled flux qubits the
ground-state and the spectroscopic measurements give the same
results.

\begin{figure}[th]
\includegraphics[width=6 cm,angle=0]{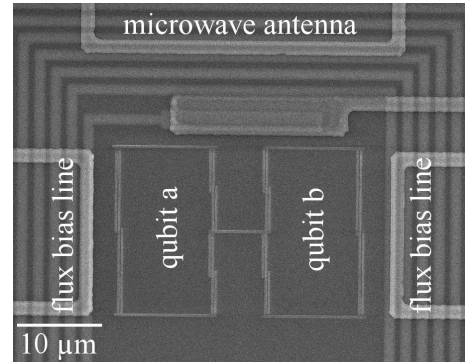}
\caption{Scanning electron micrograph showing two coupled aluminium flux
qubits, placed inside a niobium tank coil and $dc$ and microwave lines. The
qubits are fabricated on an oxidized silicon substrate by making use of
electron-beam lithography and shadow evaporation. The loops share a large
Josephson junction visible in the center. This junction provides a coupling
between the qubits with a coupling energy $J/h=1.9$ GHz. The qubit loops are
interrupted by three junctions each. Two of them, located in the inner
sides, have nominal areas of $200\times 700$ nm while the outer junctions
are 70\% smaller. The qubit system is placed inside a 30 turn
superconducting niobium pancake coil which forms a resonant tank circuit,
with a resonance frequency of 20.8 MHz, with a capacitor mounted on the
sample holder. The resonator typically has a quality $Q=300$. Individual
control of the dc-flux bias of the individual qubits ($\Phi _{a},\Phi _{b}$%
), is provided by the niobium bias lines visible on top of the coil
windings. The antenna used for supplying the MW-excitation is visible at the
top of the picture.}
\label{photo}
\end{figure}

The persistent current, or flux, qubit is a small superconducting loop with
three submicron Josephson junctions~\cite{Mooij99}. Due to the flux
quantization only two phases are independent. Thus, the circuit is
characterized by a two dimensional potential $U(\phi _{1},\phi _{2})$ which,
for suitable qubit parameters, exhibits two minima. In the classical case
these minima correspond to clockwise and anticlockwise supercurrents in the
loop. If the applied magnetic flux equals half a flux quantum, $\Phi
_{x}=\Phi _{0}/2$ ($\Phi _{0}=h/2e$), both minima have the same potential,
leading to a degenerate ground state. According to the quantum mechanics the
degeneracy is lifted close to this point and the flux qubit can be described
by the Hamiltonian \cite{Makhlin01}:
\begin{equation}
H(t)=-\Delta \sigma _{\!x}-\varepsilon \sigma _{\!z}+A\cos (\omega t)\sigma
_{\!z}\;,  \label{eq01}
\end{equation}%
where $\sigma _{x},\sigma _{z}$ are the Pauli matrices for the spin basis
and $\Delta $ is the tunneling amplitude. The qubit bias is given by $%
\varepsilon =I_{p}(\Phi _{x}-\Phi _{0}/2)$, where $I_{p}$ is the magnitude
of the qubit persistent current. The last term describes the microwave
irradiation necessary for the spectroscopy which aims to probe the
stationary energy levels represented by the first two, time independent,
terms of this Hamiltonian. The eigenvalues of the Hamiltonian (\ref{eq01})
depend on the flux bias $\Phi =\Phi _{x}-\Phi _{0}/2$:%
\begin{equation}
E_{\pm }(\Phi )=\pm \sqrt{(I_{p}\Phi )^{2}+\Delta ^{2}}.
\label{eq02}
\end{equation}%
Spectroscopic measurements detect the excitation of a qubit close to the
point where the microwave irradiation matches the level spacing: $\hbar
\omega =\Delta E(\Phi )\equiv E_{+}(\Phi )-E_{-}(\Phi )$. By measuring $%
\Delta E$ as a function of $\Phi $ the qubit parameters $\Delta $
and $I_{p}$ can be obtained as has been shown by van der Wal
\textit{et al.}~\cite{Wal00}.

Alternatively, the same information can be obtained from ground-state
measurements. Indeed, let us consider a flux qubit weakly coupled to a
classical oscillator consisting of an inductor $L_{T}$ and a capacitor $%
C_{T} $ forming a tank circuit~\cite{com}. Due to the mutual inductance $M$
the tank biases the qubit resulting in $\Phi =\Phi _{dc}+\Phi _{rf}$.
Provided that the resonant frequency of the tank is small, $\omega _{T}=1/%
\sqrt{L_{T}C_{T}}\ll \Delta /\hbar $, and the temperature is low enough, $%
k_{B}T\ll 2\Delta $ ($k_{B}$ is Boltzmann's constant), the qubit will reside
in its ground state $E_{-}$. The dynamic behavior of the tank-qubit
arrangement can be described by the Lagrangian
\begin{equation}
\mathcal{L=T-U}=\frac{1}{2}L_{T}\dot{q}^{2}+E_{-}(\Phi _{dc}+M\dot{q})-\frac{%
1}{2}\frac{q^{2}}{C_{T}},  \label{Eq:L}
\end{equation}%
where $q$ is the charge on the tank capacitor and $\dot{q}$ is the
circulating current in the tank. Such Lagrangian would lead to the nonlinear
equation of motion:
\begin{equation}
0=\frac{d}{dt}\frac{\partial \mathcal{L}}{\partial \dot{q}}-\frac{\partial
\mathcal{L}}{\partial q} \\
=\left( L_{T}+M^{2}\frac{\partial ^{2}E_{-}(\Phi _{dc}+M\dot{q})}{\partial
\Phi ^{2}}\right) \ddot{q}+\frac{q}{C_{T}},  \label{Eq:NonLin}
\end{equation}%
however for small amplitude of $\dot{q}$ the Lagrangian can be linearized~%
\cite{note} by replacing $E_{-}(\dot{q})$ by its second order Taylor
expansion around $\Phi _{dc}$. Consequently, we obtain the simple Lagrangian
of a particle in a parabolic potential well:
\begin{equation}
\mathcal{L}=\frac{1}{2}m^{\ast }\dot{q}^{2}-\frac{1}{2}k^{\ast }q^{2}.
\label{Eq:Lfin}
\end{equation}%
The equation of motion which can be obtained from this Lagrangian is just the simple
equation for a particle in a parabolic potential, $m^{\ast }\ddot{q}=k^{\ast
}q$, where
\begin{equation}
m^{\ast }=\left( L_{T}+M^{2}\frac{d^{2}E_{-}(\Phi _{dc})}{d\Phi _{dc}^{2}}%
\right) ,  \label{Eq:m}
\end{equation}%
is the effective mass and $k^{\ast }=1/C_{T}$ is the curvature of the
parabolic potential well. Thus, the resonant frequency of the tank-qubit
arrangement
\begin{equation}
\omega _{0}=\sqrt{\frac{k^{\ast }}{m^{\ast }}}\approx \omega _{T}\left( 1-%
\frac{M^{2}}{2L_{T}}\frac{d^{2}E_{-}(\Phi _{dc})}{d\Phi _{dc}^{2}}\right) ,
\label{EQ:meff}
\end{equation}%
contains information on the curvature of the ground state of the qubit.
Differentiating Eq. (\ref{eq02}) results in:
\begin{equation}
\frac{d^{2}E_{-}(\Phi _{dc})}{d\Phi _{dc}^{2}}=-\frac{(I_{p}\Delta )^{2}}{%
\left( \varepsilon ^{2}(\Phi _{dc})+\Delta ^{2}\right) ^{3/2}}\;,
\label{Eq:dE}
\end{equation}%
showing that $\Delta $ and $I_{p}$ can be determined from the dependence of
the resonance frequency of the tank circuit on the applied flux.

\begin{figure}[thb]
\includegraphics [scale=0.5] {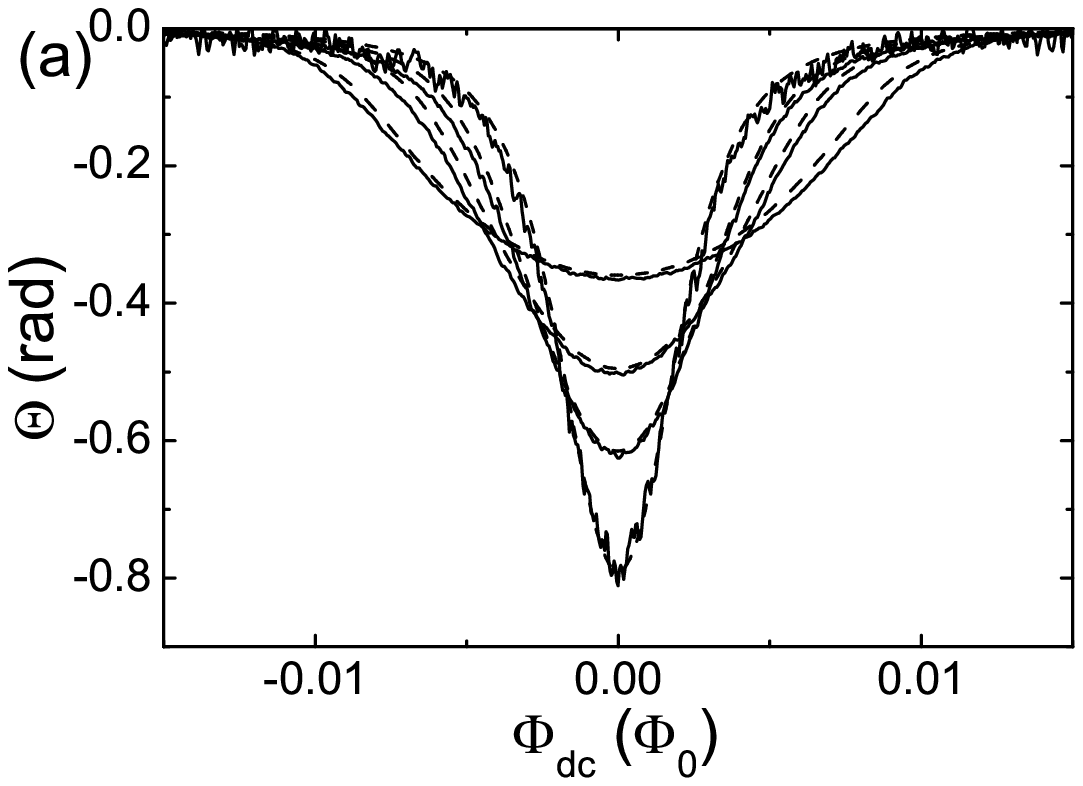}
\includegraphics [scale=0.5]
{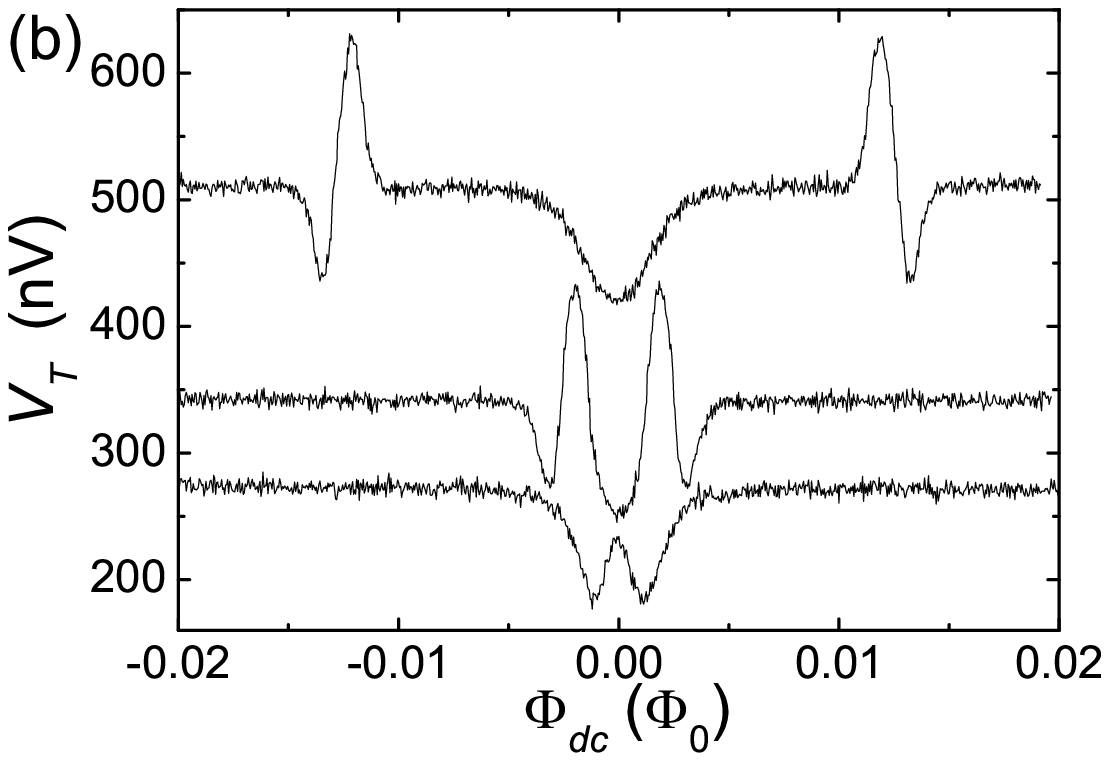} \includegraphics [scale=0.5] {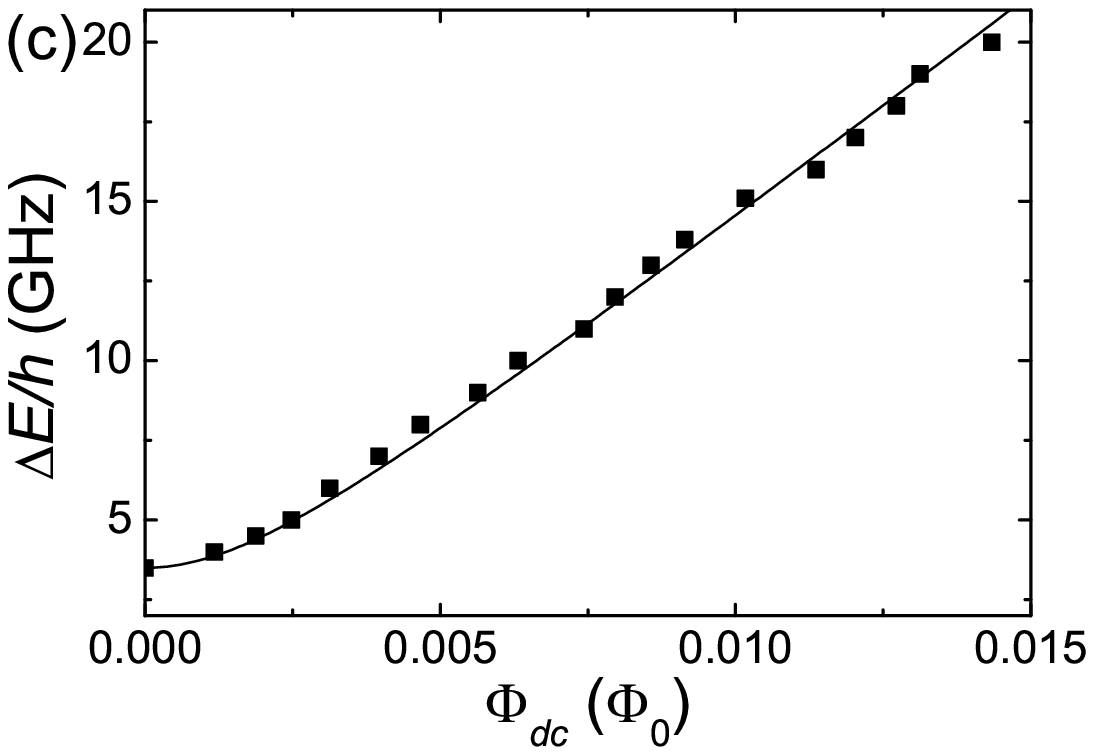}
\caption{Comparison of the ground state and spectroscopic measurements for
qubit $b$. (a) Ground state measurements. Presented is the dependence of the
phase shift between the tank circuit voltage and bias current on the flux
bias. The solid lines are experimental data fitted by the theoretical curves
(dashed curves) for qubit parameters $I_{p}=225$~nA and $\Delta /h=1.75$
GHz. The curves correspond to various values of the $rf$-bias current on the
tank circuit resulting in $rf$ voltage amplitudes, from top to bottom, $%
V_{T}=4.3,2.9,0.5,$ and $0.3~\protect\mu $V. (b) Amplitude of the tank
voltage as a function of the normalized magnetic flux in the qubit at the
driving frequencies, from top to bottom, $\protect\omega /2\protect\pi =18$,
$5$, and $3.5$ GHz. The curves have been shifted for clarity. The resonant
excitation in the flux qubit results in the peak-and-dip at the positions
defined by the condition $\Delta E(\Phi _{dc})=\hbar \protect\omega $. (c)
Energy gap $\Delta E$ between the qubit energy levels determined from the
positions of the mid points of the peak-and-dip structures (solid squares).
The solid line is the theoretical curve calculated from Eq.~(\protect\ref%
{eq02}) using the parameters $I_{p}=225$~nA and $\Delta /h=1.75$ GHz
obtained from the ground-state measurements. The effective temperature $%
T\approx 70$ mK$\approx 1.4$ GHz$\cdot h$/$k_{B}$ is smaller than the
minimal energy level separation $2\Delta $.}
\label{Fig:OneQubit}
\end{figure}

In order to compare both methods we fabricated a two qubit sample like the
one shown in (Fig.~1). As either one of the qubits can be biased far away
from degeneracy, the single qubit properties can be studied as well. This
can be understood if we consider the Hamiltonian of two coupled flux qubits:
\begin{equation}
H_{2qbs}=-\Delta _{a}\sigma _{x}^{(a)}-\Delta _{b}\sigma
_{x}^{(b)}-\varepsilon _{a}\sigma _{z}^{(a)}-\varepsilon _{b}\sigma
_{z}^{(b)}+J\sigma _{z}^{(a)}\sigma _{z}^{(b)},  \label{Ham_2qbs}
\end{equation}%
where $J$ is the Josephson coupling energy provided by the large connecting
Josephson junction. Suppose qubit $a$ is the one biased far from its
degeneracy point in such a way that $\varepsilon _{a}$ is large in
comparison with the other energy variables. Then, qubit $a$ has a well
defined ground state with averaged spin variables $\left\langle \sigma
_{z}^{(a)}\right\rangle =1$ and $\left\langle \sigma _{x}^{(a)}\right\rangle
=0$ which can be averaged out of the two-qubit Hamiltonian (\ref{Ham_2qbs})
reducing it to: $H_{2qbs,red}=-\Delta _{b}\sigma _{x}^{(b)}-(\varepsilon
_{b}-J)\sigma _{z}^{(b)}$. Apart from the offset in the bias term due to the
coupling this is identical to the single qubit Hamiltonian (\ref{eq01}).
This offset can be easily compensated and measured allowing the
determination of the coupling energy $J$ \cite{Grajcar05}. The qubit
parameters, $\Delta _{b}$ and $I_{p}^{(b)}$, are determined from the ground
state measurement, as it is described above. Analogously, biasing qubit $b$
far from the degeneracy point the parameters for qubit $a$, $\Delta _{a}$
and $I_{p}^{(a)}$, can be determined. In a similar way the parameters of a
$N $-qubit Hamiltonian can be completely reconstructed from the ground-state
measurements as has already been demonstrated, for instance, for four qubits
circuits \cite{Grajcar06}.

The qubit parameters can be probed by either the ground state (adiabatic)
measurements or by making use of spectroscopy. However it naturally raises
the question of whether the ground-state and the spectroscopy measurements
are consistent? While addressing this problem in this Letter we study both
approaches \textit{in situ} for one and two coupled flux qubits.

Experimentally, the shift of the resonance frequency can be obtained by
driving the tank circuit with a $rf$ current $I_{rf}$ at a frequency close
to the resonant frequency $\omega _{T}$ and measure the phase shift $\Theta $
between the $rf$ voltage and driving current. For a small qubit inductance~$L
$, the phase shift $\Theta $ is defined by~\cite{Greenberg02b}:
\begin{equation}
\tan \Theta =\frac{M^{2}Q}{L_{T}}\frac{d^{2}E_{-}}{d\Phi _{dc}^{2}}.
\end{equation}%
The mutual inductance $M$, tank inductance $L_{T}$ and quality
factor $Q$ can be measured independently giving a value of $23.4$ pH
for this prefactor. The results of such measurements are shown in
Fig.~2(a). Note that the sample was thermally anchored to the mixing
chamber of a dilution refrigerator at a temperature $T_{mix}\approx
10$ mK. The effective temperature of the sample $T$ is higher and we
estimated from the best theoretical fits that $T\approx 70$ mK
\cite{our}.

It is important to note that thermal excitations can
modify the measured signal, which would result in erroneous qubit parameters.
In practice, thermal excitations are not negligible when $k_{B}T \gtrsim 2\Delta$.
Nevertheless, if $k_{B}T < 2\Delta $  the dispersive measurement provides a correct value of
qubit parameters [11]. This statement is also confirmed  by a good agreement
between both ground state and spectroscopic measurements (see Fig. 2(c)).

 \begin{figure}[ht]
\includegraphics[scale=1]{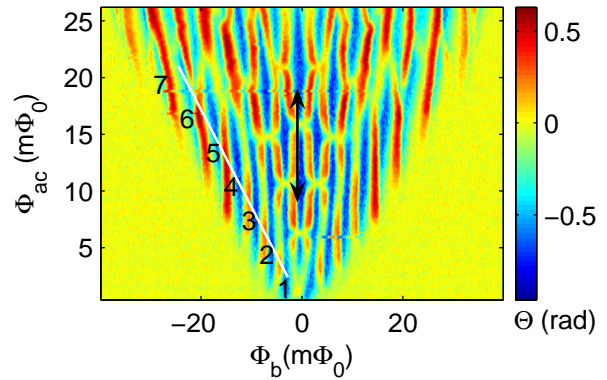}
\caption{(color online). Landau-Zener interferometry for qubit $b$.
Dependence of the tank voltage phase shift $\Theta $ on the \textit{dc} flux
bias $\Phi _{dc}$ and the \textit{ac} flux amplitude $\Phi _{ac}$ (the
microwave amplitude). The spots along the $\Phi _{dc}$ axis correspond to
the multiphoton resonances at the positions defined by the relation $\Delta
E(\Phi _{dc})\approx n\cdot \hbar \protect\omega $; the numbers from $1$ to $%
7$ show the position of the $n$-photon resonances. The changes along the $%
\Phi _{ac}$ axis are due to St\"{u}ckelberg oscillations in the qubit. The
calibration of the driving power of the $ac$ flux can be done either with
the distance between these oscillations (black arrow) or from the slope of
the interference fringes (white line).}
\label{LZI}
\end{figure}

In fact the tank circuit can be used as detector for the spectroscopy
measurements as well, since the variation in the population of the qubits'
energy levels results in the change of the effective impedance of the tank
circuit \cite{Grajcar07}. The tank circuit is insensitive to the microwave
signal itself since $\omega _{T}\ll \Delta /\hbar $ and $Q\gg 1$. However,
if the microwave frequency is close to the qubit level separation, the
system damps or amplifies the voltage on the tank, mimicking the Sisyphus
mechanism of damping (and heating) of the tank known from quantum optics
\cite{Wineland92}. This effect generates the peak-dip structure in the $V{%
_{T}}(\Phi _{dc})$ dependence around the resonance (see Fig.~2(b)) \cite%
{Grajcar07, Shevchenko07}. The position of the resonances is the point where
a peak changes to a dip. From the positions of the mid points of the
peak-dip structures one can determine the energy gap $\Delta E$ between the
energy levels. The obtained agreement between the adiabatic and
spectroscopic measurement for weak driving regime is excellent (see
Fig.~2(c)).

With increasing the microwave power, the Landau-Zener interference
pattern of the qubit is clearly visible. The qubit's response in the
strong driving regime is demonstrated in Fig.~3 where the tank
voltage phase shift is presented as a function of the microwave
amplitude and the $dc$ flux bias. The position of the multiphoton
resonances is approximately given by the relation $\Delta E(\Phi
_{dc})\approx n\cdot \hbar \omega $, where the energy gap $\Delta E$
is calculated using the parameters $I_{p}$ and $\Delta $ obtained
from the ground state measurement. Moreover, the Landau-Zener
interferometry allows the calibration of the microwave power to the
$ac$ flux due to the periodicity of the St\"{u}ckelberg oscillations on
the
parameter ${4I_{p}\Phi _{ac}}/{\hbar \omega }$ with the period $2\pi $ \cite%
{Shytov2003, Shevchenko2006}. It follows that the distance between the
resonances (shown by the black arrow in Fig.~3) is approximately equal to $%
\delta \Phi _{ac}={\textstyle\frac{1}{2}}{\pi \hbar \omega }/{I_{p}}$.
Alternatively, the calibration can be made using the slope of the
interference fringes (white line in Fig.~3) \cite{Oliver05, Sillanpaa06a}.

\begin{figure*}[th]
\includegraphics[scale=0.9]{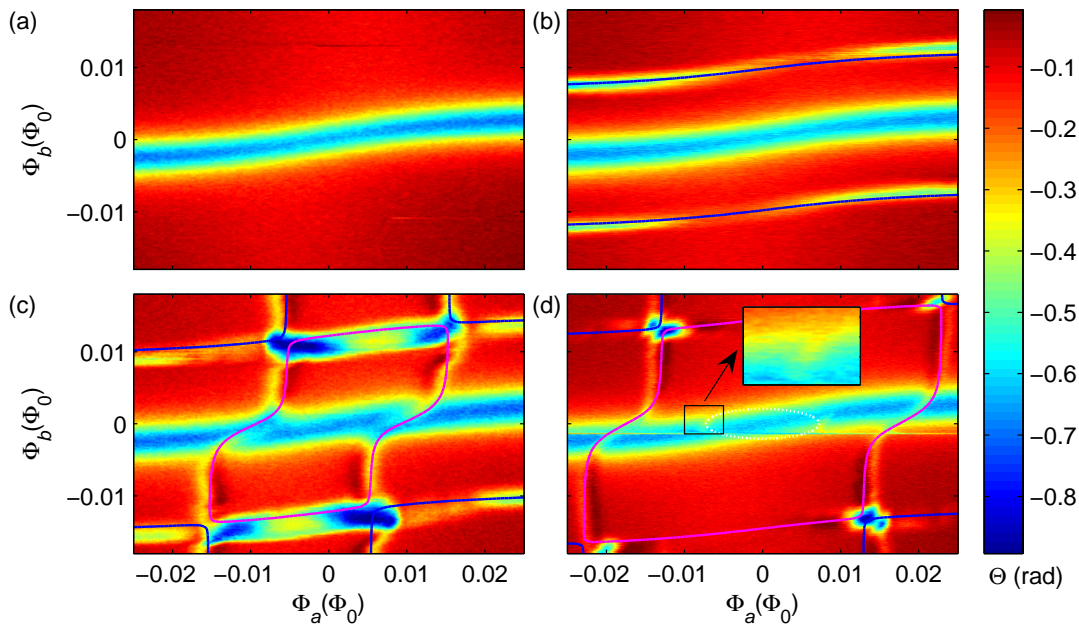}
\caption{(color online). Spectroscopy of the system of two coupled flux
qubits. The dependence of the tank voltage phase shift $\Theta $ on the flux
biases in qubits $a$ and $b$ is presented for measurements without microwave
excitation in (a) and for microwave excitation with $\protect\omega /2%
\protect\pi =$~14.125, 17.625 and 20.75 GHz in (b) till (d) respectively.
Inset shows the transition to the third excited level. The blue, magenta and
white-dotted lines in the pictures with microwave excitation show the
expected positions of the resonant excitations of the qubits to the first,
second and third excited levels respectively, calculated from the energy
eigenvalues of Hamiltonian (\protect\ref{Ham_2qbs}) with parameters: $\Delta
_{a}/h=7.9$ GHz, $\Delta _{b}/h=1.75$ GHz, $I_{p}^{(a)}=120$ nA, $%
I_{p}^{(b)}=225$ nA, and $J/h=1.9$ GHz. The trough around $\Phi _{b}=0$ is
due to the ground state curvature of qubit $a$ and corresponds to the ground
state measurements of Fig. (\protect\ref{Fig:OneQubit}). The shallow trough
around $\Phi _{a}=0$ visible in figure (a), is due to qubit $a$.}
\end{figure*}

After determining the single qubit parameters far away from the degeneracy
points, we investigated the two qubit behavior. Firstly, the coupling energy
$J$ was determined from the offset of the qubit dips from the $\Phi _{a/b}=0$
lines, visible in the pure ground state measurements presented in Fig.~4(a).
Then the qubits were driven by various $ac$ magnetic fluxes $\Phi _{ac}\sin
\omega t$. In Fig.~4(b) a frequency in-between both qubit gaps was used and
therefore only the transitions to the first excited state are visible. For
higher frequencies, also the second and third excited states become visible
as can be seen in subfigures (c) and (d). Here also both types of the
measurements (ground-state and spectroscopic) result in the same set of
parameters for the system. Finally we would like to note that the
theoretical calculations allow us to plot analogous to Figs. 3 and 4 graphs
(to be published elsewhere \cite{our}).

In conclusion, the equivalence of the ground-state and spectroscopic
approaches for the measurement of the qubit system parameters was
demonstrated. We have probed the one- and two- flux qubit systems by using a
dispersive measurement technique. It was shown that the ground state
measurement gives the same qubit parameters as the spectroscopy in the weak
(Figs.~2 and 4) as well as in the strong driving regime (Fig.~3).

\begin{acknowledgements}
This work was initiated by A. Izmalkov who untimely passed away when this
work was finished. We thank Yakov Greenberg for useful discussions and
gratefully acknowledge the financial support of the EU through the RSFQubit
and EuroSQIP projects. M. G. was supported by Grants APVV-0432-07 and VEGA
1/0096/08. S. N. S. acknowledges the financial support of INTAS through a YS
Fellowship Grant.
\end{acknowledgements}




\begin{thebibliography}{99}
\bibitem{com} This arrangement has been already described phenomenologically
-- oscillator perturbed by two-level system \cite{Greenberg02b}. Here we
present a more general approach to this problem.
\bibitem{note} For finite $I_{rf}$ the response of the tank circuit can be
calculataed numerically \cite{Greenberg02b}.
\end{thebibliography}
\end{document}